\documentclass[preprint,aps]{revtex4}

\usepackage{graphicx}
\usepackage{dcolumn}
\usepackage{bm}

\begin{document}
\preprint{}
\title{Nucleon bound state at finite temperature}
\author{I. Zakout}
\affiliation{Institut f\"ur Theoretische Physik, 
J. W. Goethe Universita\"at, Robert-Mayer-Stra$\beta$e 8-10,
Postfach 11 19 32, D-60054 Frankfurt am main, Germany}


\begin{abstract}
We study the nucleon at finite temperature using Bethe-Salpeter equation (BSE) 
where it is considered as a bound state of a scalar-diquark and a quark. 
The interaction between the diquark and the quark is taken as an exchange quark. 
The constituent quark and diquark substructures are considered 
momentum and temperature dependent based on the lattice QCD 
and dynamical chiral symmetry breaking (DCSB). 
The kernel interaction modification with temperature is 
approximated using the mixed representation in the imaginary time formalism 
and then we adopt an adiabatic (instantaneous) approximation 
after evaluating the Matsubara frequencies sum.
The results of BSE are compared with those for MIT bag model. 
We argue the bag pressure is a temperature dependent and it decreases 
smoothly with temperature to be consistent with the results for BSE. 
Furthermore, the nucleon bound state obtained by BSE ceases to exist 
just below the critical temperature and this indicates that the nucleon 
deconfines to its constituent quarks at the critical temperature. 
The nucleon mass decreases with temperature 
as  $M_N(T)/M_N(0)\approx {\cal R}^{\delta}(T)$ 
where ${\cal R}(T)=\left(1-b\frac{T^2}{T^2_c}\right)^{1/2}$ 
and $\delta\approx 1$ 
and this suggests $B(T)/B(0)={\cal R}^{4\delta}(T)$ for MIT bag. 
\end{abstract}

\maketitle
\section{Introduction}

The properties of hadrons and their masses at finite density and temperature 
are  extremely important to study the phase transition from 
the hadronic phase to the quark gluon plasma.
Generally speaking, the nonperturbative calculations at finite 
temperature and/or density support the idea that
hadronic masses change as a function of temperature and density.
The nucleon mass is found to depend substantially on temperature variations 
of the quark condensate and interaction process. 
As the condensate disappears smoothly with chiral-symmetry restoration 
so does the nucleon 
mass\cite{Adami:1990sv,Adami:1992at,Kacir:1995gy,Ropke:1986mh}.

Recently, baryons have been studied in the context of Bethe-Salpeter 
equation (BSE). The baryon is considered as a bound state of the confined 
constituent diquark and quark interacting via an exchange quark.
Despite the simplicity of the model, it has been used to study the intermediate
energy physics successfully
\cite{Keiner:1995bu,Keiner:1996at,Hellstern:1997pg,Oettel:2000jj,
Alkofer:2004yf}. 
Furthermore, it has been some attempts to extend BSE to study mesons 
at finite temperature\cite{Blaschke:1998gk,Blaschke:2000gd,Roberts:2000aa}.
However, the dressed quark propagator and the variation of 
the constituent quark mass with temperature are essential to study 
the bound states in a hot bath. 
The Schwinger-Dyson equation (DSE) is used to study the quark propagator at finite
temperature\cite{Roberts:2000aa}. 
The DSE with QCD potential can be solved numerically at finite temperature
\cite{Ikeda:2001vc}. 
Furthermore, the constituent quark mass and its variation with temperature 
is studied using the lattice-QCD-based Schwinger-Dyson equation. 
Spontaneously broken chiral symmetry is found to be 
restored at the critical temperature\cite{Iida:2003gq,Iida:2004ih}.
The linear confinement dissociates smoothly with 
temperature\cite{Pisarski:1982cn}. 
From DSE, it is expected that 
the dressed quark mass to decrease with temperature 
in the same order of the linear confinement dissociation.  
The condensate is found also to disappear smoothly at the same order.

In Sec. II, we shall derive the BSE for a quark-diquark bound state at finite temperature.
The modification of the interaction with temperature is approximated using 
a mixed representation of the imaginary time formalism 
and then we adopt the adiabatic (instantaneous) approximation.
In this approximation, we set the time component to zero after evaluating 
the Matsubara sums not before. 
The constituent scalar-diquark and quark substructures are effectively taken into account 
and their masses are taken momentum and temperature dependent based on the lattice-QCD.
In Sec. III, we briefly review the MIT bag model for nucleon 
and then we extend it to finite temperature case
by taking on the considerations the dissociation 
of the nucleon bag pressure with temperature.
Finally in Sec. V, we present our results and conclusions.

\section{Diquark-quark Bethe-Salpeter for Nucleon}
The baryon is considered as a bound state of confined
constituent diquark and quark interacting via quark exchange.
The BSE for the diquark and quark bound states is given by
\begin{eqnarray}
\chi_P(p)=
\Delta_{D}(p_1)S_q(p_2) \int \frac{d^4 p'}{(2\pi)^4}
\left[-iK(P,p,p')\right]\chi_P(p'),
\end{eqnarray}
where 
\begin{eqnarray}
p_1=\eta_1 P + p \nonumber \\
p_2=\eta_2 P - p.
\end{eqnarray}
We have taken $K(P,p,p')=\phi(q')S_q(q)$ 
where $\phi(q)$ is the quark-diquark vertex coupling $g^2_{Dq}\equiv g^2_{Dq}(q)$ 
and $S_q(q)$ is the exchange quark propagator and $q=p+p'$. 
The scalar diquark propagator reads
\begin{eqnarray}
\Delta_{D}(p_1)&=&\left(\frac{1}{Z_D}\right) 
\frac{i}{[p^2_1-m^2_1+i\delta]} \nonumber \\
&=&\left(\frac{1}{Z_D}\right) 
\frac{i}{2\epsilon_1}\left[
\frac{1}{p_0+M_1-\epsilon_1+i\delta}
-
\frac{1}{p_0+M_1+\epsilon_1-i\delta}
\right],
\end{eqnarray}
and the fermion quark propagator reads
\begin{eqnarray}
S_q(p_2)&=&
i \left(\frac{1}{Z_q}\right)\frac{\gamma\cdot p_2+m_2}{p^2_2-m^2_2+i\delta} 
\nonumber\\
&=&-i\left(\frac{1}{Z_q}\right)
\left[
\frac{\Lambda_{+}(-{\bf p})}{p_0-M_2+\epsilon_2-i\delta}
+
\frac{\Lambda_{-}(-{\bf p})}{p_0-M_2-\epsilon_2+i\delta}
\right]\beta,
\end{eqnarray}
where $\epsilon_i=\sqrt{{\bf p}^2+m^2_i}$ and $i=1$ refers to the scalar diquark 
while $i=2$ refers to the quark and $M_1=\eta_1 M$ and $M_2=(1-\eta_1) M$.
We have chosen the proper choice $\eta_1=m_1/(m_1+m_2)$. 
In the center of mass frame we have $P=(M,0)$.
The fermion positive and negative energy projections read
\begin{eqnarray}
\Lambda_{a}(-{\bf p})=\frac{\epsilon_2+a\beta(-\gamma\cdot{\bf p}+m_2)}{2\epsilon_2},
\end{eqnarray}
where $a=\pm$.
It is possible to take $Z_D=1$ and $Z_q=1$ with a good approximation. 
In principle, these functions can be considered in the calculations 
by adopting smearing functions in the propagators\cite{Hellstern:1997pg,Oettel:2000jj}. 
In the present calculations, we shall assume these functions are absorbed 
by the quark-diquark vertex coupling $g^2_{Dq}$.   
The constituent diquark and quark masses are given by 
$m_D=m_D(p)$ and $m_q=m_q(p)$ for the diquark and quark, respectively.
In the adiabatic approximation, the temperature response is separated
from the momentum dependent and the constituent diquark and quark masses 
reduce to $m_D\equiv m_D({\bf p};T)$ and $m_q\equiv m_q({\bf p};T)$.
The Dynamical chiral-symmetry breaking (DCSB) is one of the most nonperturbative QCD feature.
Based on the Lattice QCD at the quenched level and Landau gauge 
and the lattice data in the chiral limit, 
the quark mass function at zero temperature is reproduced by
\begin{eqnarray}
m_{u,d}(p^2)=m_0+\frac{m_{Q}(T=0)}{\left[ 1+(p^2_E/\overline{p}^2)^{\gamma/2}\right]},
\label{qconstit}
\end{eqnarray}
where $\overline{p}=$ 870 MeV, $\gamma=3.04$, the current mass $m_0=0$ MeV
and $p^2_E=p^2_4+{\bf p}^2$ is the momentum in Euclidean space.
Using the lattice-QCD-based gap equation (DSE) at finite temperature, it is
found that $m_Q(T)$ has a chiral symmetry restoration phase
transition at the critical temperature $T_c$. 
The quark and diquark acquire their constituent masses from the linear confinement. 
These masses can be computed using DSE to calculate the self-energy correction.
However the linear confinement constant $V(r)\approx a_{\mbox{L}}(T) r$ in the configuration space
is found to dissociate thermally spontaneously to $a_{\mbox{L}}(T)/a_{\mbox{L}}(0)={\cal R}(T)$.
The linear confining is a nonperturbative effect and it is believed to generate the major contribution
to the self-energy correction for the current quark mass to acquire its constituent value.
It is reasonable to take the self-energy correction and mass variation 
with respect to the temperature 
at the same order of the spontaneous linear confinement dissociation  
\begin{eqnarray}
m_Q(T)/m_Q(0)={\cal R}(T),
\end{eqnarray}
where ${\cal R}(T)=\sqrt{1-b T^2/T^2_c}$ for $T\le T_c$ 
and ${\cal R}(T)=\sqrt{1-b}$ for $T>T_c$\cite{Pisarski:1982cn,Kaczmarek:1999mm}.
Since the constituent diquark is assumed as a weekly quark-quark bound state, 
it is reasonable to assume 
$m_{\mbox{D}}(T)/m_{\mbox{D}}(0)={\cal R}(T)$.  
Furthermore, we assume the diquark substructure takes the same quark formula
substructure given in Eq.(\ref{qconstit}). 
In the adiabatic approximation, euclidean momentum $p^2_E=p^2_4+{\bf p}^2$
in the quark substructure is reduced to three dimensional momentum where $p^2_4=0$.
However, in the imaginary time formalism, we can write $p_4=(2n+1)\pi T$ where $n=1,2,\cdots$
and we can take $p^2_E=\pi^2 T^2+{\bf p}^2$ to the first order approximation. 
To be consistent with the adiabatic approximation considered in our calculations, 
we have adopted $p^2_4=0$ in our calculations. 
Furthermore, we have shown that the first order approximation effect is small.    

In an adiabatic approximation, the bound state equation at finite temperature
is derived as follows
\begin{eqnarray}
\int \frac{d p_0}{(2\pi)}\chi_P(p)=\left[i\int \frac{d p_0}{(2\pi)}\Delta_{D}(p_1)S_q(p_2)\right]
\int \frac{d^3 p'}{(2\pi)^3}
\left[-K(q_0,{\bf q};T)|_{q_0=0}\int \frac{d p'_0}{(2\pi)}\chi_P(p')\right].
\end{eqnarray}
At zero temperature, the adiabatic approximation is reduced 
to the instantaneous approximation. 
However, as we shall show below, a special attention must be paid for 
the interaction potential in the adiabatic approximation.
The kernel interaction is taken as an exchange quark between 
the diquark and quark,
\begin{eqnarray}
-iK(P;p,p')\chi_P(p')=-iK(q)\chi_P(p')=g^2_{Dq}S_q(p+p')\chi_P(p'),
\end{eqnarray}
where $q=p+p'$. 
The BSE propagator is reduced to
\begin{eqnarray}
{\cal G}(M,{\bf p};T)&=& i\int \frac{d p_0}{2\pi}\Delta_D(p_1)S_q(p_2)=   
\nonumber \\
&=&
{\cal G}_{+}(M,{\bf p};T) \Lambda_{+}(-{\bf p})\beta
+
{\cal G}_{-}(M,{\bf p};T) \Lambda_{-}(-{\bf p})\beta,
\label{BS_PROPG1}
\end{eqnarray}
where
\begin{eqnarray}
{\cal G}_{+}(M,{\bf p};T)
&=&
\frac{1}{2\epsilon_1}\left[
\frac{1}{M-(\epsilon_1+\epsilon_2)}
\right]\left[1+n_B\left(\epsilon_{-1}\right)
-n_F\left(\epsilon_{-2}\right)\right] \nonumber \\
&+&
\frac{1}{2\epsilon_1}\left[
\frac{1}{M+(\epsilon_1-\epsilon_2)}
\right]
\left[n_B\left(\epsilon_{+1}\right)
+n_F\left(\epsilon_{-2}\right)\right]  \nonumber \\
\end{eqnarray}
and
\begin{eqnarray}
{\cal G}_{-}(M,{\bf p};T)
&=&
\frac{1}{2\epsilon_1}\left[
\frac{1}{M+(\epsilon_1+\epsilon_2)}
\right]
\left[1+n_B\left(\epsilon_{+1}\right)
-n_F\left(\epsilon_{+2}\right)\right]  \nonumber \\
&+&
\frac{1}{2\epsilon_1}\left[
\frac{1}{M-(\epsilon_1-\epsilon_2)}\right]
\left[n_B\left(\epsilon_{-1}\right)
+n_F\left(\epsilon_{+2}\right)\right],
\end{eqnarray}
where $\epsilon_{\pm i}=\epsilon_i \pm M_i$.
The thermal distribution functions for quark and scalar diquark
$n_F(x)=\frac{1}{e^{x/T}+1}$ and $n_B(x)=\frac{1}{e^{x/T}-1}$
are the Fermi-Einstein and Bose-Einstein distribution functions, respectively, 
and $x\ge 0$.
At zero temperature, the BSE propagator given by Eq.(\ref{BS_PROPG1}) is reduced to
\begin{eqnarray}
i\int \frac{d p_0}{2\pi}\Delta_1(p_1)S_2(p_2)|_{T=0}=
\frac{1}{2\epsilon_1}
\left[\frac{\Lambda_+(-p)\beta}{+M-(\epsilon_1+\epsilon_2)}
+\frac{\Lambda_-(-p)\beta}{+M+(\epsilon_1+\epsilon_2)}\right].
\end{eqnarray}
It is a wonderful to write the thermal distribution function of BSE propagator 
in a proper way to discuss thermal reaction processes.   
The thermal distribution function for the positive energy solution is written as follows
\begin{eqnarray}
\left[1+n_B(\epsilon_{-1})-n_F(\epsilon_{-2})\right]=
\left[1+n_B(\epsilon_{-1})\right]\left[1-n_F(\epsilon_{-2})\right]
+n_B(\epsilon_{-1})n_F(\epsilon_{-2}).
\end{eqnarray}
It corresponds to the difference between the virtual decay and creation rates 
via the virtual reaction processes 
$\phi_M\rightarrow\phi_1\phi_2$ and $\phi_1\phi_2\rightarrow\phi_M$, respectively. 
It is interested to note that the constituent diquark and quark chemical potentials
$M_1$ and $M_2$, respectively, 
are always less than their relativistic energies 
and $\epsilon_{-1}=\epsilon_{1}-M_1>0$ and $\epsilon_{-2}=\epsilon_{2}-M_2>0$
to preserve the confinement condition. 
This guarantees the creation rate is always larger than the decay rate. 
Furthermore, the additional `Landau Damping' term 
\begin{eqnarray}
\left[n_B(\epsilon_{+1})+n_F(\epsilon_{-2})\right]
=
n_B(\epsilon_{+1})\left[1-n_F(\epsilon_{-2})\right]
+\left[1+n_B(\epsilon_{+1})\right]n_F(\epsilon_{-2}),
\end{eqnarray}
corresponds the virtual decay and creation processes via the reaction scattering
$\overline{\phi}_1\phi_M\rightarrow\phi_2$ and $\phi_2\rightarrow\overline{\phi}_1\phi_M$, respectively. 
This corresponds particles disappear or are created through the scattering in the bath, and not via 
the process which are available at zero temperature.
The thermal distribution function for the negative energy solution 
corresponds to 
$\overline{\phi}_M\rightarrow\overline{\phi}_1\overline{\phi}_2$
and
$\overline{\phi}_1\overline{\phi}_2\rightarrow\overline{\phi}_M$ 
for the decay and creation processes, respectively.
On the other hand, the `Landau Damping' term in the negative energy solution 
reduces to the creation and decay processes 
$\overline{\phi}_2\rightarrow\phi_1\overline{\phi}_M$ and
$\phi_1\overline{\phi}_M\rightarrow\overline{\phi}_2$, respectively.

The exchange quark propagator is decomposed as follows
\begin{eqnarray}
S_q(q)=i\beta
\left[
\frac{\Lambda_{+}(-{\bf q})}{q_0-\epsilon_q}
+
\frac{\Lambda_{-}(-{\bf q})}{q_0+\epsilon_q}
\right].
\end{eqnarray}
Hence the kernel interaction reads
\begin{eqnarray}
-K(q_0,{\bf q})=
g^2_{Dq}\beta\left[
\frac{\Lambda_{+}(-{\bf q})}{q_0-\epsilon_q}
+
\frac{\Lambda_{-}(-{\bf q})}{q_0+\epsilon_q}
\right].
\end{eqnarray}
When we substitute $q_0=0$, we get
\begin{eqnarray}
-K(q_0=0,{\bf q})=
-g^2_{Dq}\beta\left[
\frac{\Lambda_{+}(-{\bf q})-\Lambda_{-}(-{\bf q})}{\epsilon_q}
\right].
\end{eqnarray}
However, this choice is not adequate in the adiabatic approximation.
In the adiabatic assumption, we shall consider the instantaneous approximation
after evaluating the Matsubara frequencies sum 
in the mixed representation for the imaginary time formalism.
The Fourier transformation of the kernel interaction 
to the mixed representation reduces to
\begin{eqnarray}
-K(\tau,{\bf q};T)&=&T\sum_n e^{-q_0\tau} K(q_0,{\bf q})\\
&=& -g^2_{Dq}
\left[
e^{-\epsilon_q\tau}
\left[1-n_F\left(\epsilon_q\right)\right]
\beta\Lambda_{+}(-{\bf q})
+
e^{\epsilon_q\tau}
\left[n_F\left(\epsilon_q\right)\right]
\beta\Lambda_{-}(-{\bf q})
\right]. \nonumber \\
\end{eqnarray}
The inverse transformation of the mixed representation 
with an adiabatic approximation reads
\begin{eqnarray}
-K(q_0,{\bf q};T)|_{q_0=0}
&=&
\int^{\frac{1}{T}}_0
d\tau e^{q_0\tau}|_{q_0=0}K(\tau,{\bf q};T) \nonumber\\
&=&
-g^2_{Dq}\frac{1}{\epsilon_q}\left[1-2n_F\left(\epsilon_q\right)\right]
\beta\left[\Lambda_{+}(-{\bf q})+\Lambda_{-}(-{\bf q})\right],
\nonumber\\
&=&
-g^2_{Dq}\frac{1}{\epsilon_q}\left[1-2n_F\left(\epsilon_q\right)\right]\beta,
\end{eqnarray}
where $\epsilon_q=\sqrt{({\bf p}+{\bf p'})^2+m^2_q}$.
It is worth to note here that in the adiabatic approximation 
we have substituted $q_0=0$ in the mixed representation 
after evaluating the Matsubara frequencies sum.
We have introduced in BSE the following interaction potential
\begin{eqnarray}
-K(q_0,{\bf q};T)|_{q_0=0} =V({\bf q};T)\beta,
\end{eqnarray}
where $V({\bf q};T)=-g^2_{Dq}\frac{1}{\epsilon_q}
\left[1-2n_F\left(\epsilon_q\right)\right]$.
The BSE reduces to
\begin{eqnarray}
\lambda\phi({\bf p})|_{\lambda=1} =
{\cal G}(M,{\bf p};T)\beta\int\frac{d^3{\bf p'}}{(2\pi)^3}
V({\bf q};T)\phi({\bf p'}).
\end{eqnarray}
The BSE wavefunction decomposes 
as Dirac wavefunction for positive parity 
\begin{eqnarray}
\phi^{l}_{jm}=
\left(
\begin{array}{c}
\eta_{lj}({\bf p})\\
\sigma\cdot\hat{\bf p}\zeta_{lj}({\bf p})
\end{array}\right)\chi^l_{jm}.
\end{eqnarray}
The BSE wavefunction reads
\begin{eqnarray}
\phi({\bf p})=
\left(
\begin{array}{c}
\phi^{+}({\bf p})\\
\phi^{-}({\bf p})
\end{array}\right)
=
\left(
\begin{array}{c}
\eta({\bf p})\\
\sigma\cdot\hat{\bf p}\zeta({\bf p})
\end{array}\right)\chi,
\end{eqnarray}
where $\chi$ is the spinor and the quantum numbers are suppressed.
The decompositions of the BSE wave function with respect 
to the positive and negative energy components read
\begin{eqnarray}
\Lambda_{+}(-{\bf p})\phi({\bf p'})=\frac{1}{2\epsilon_2}
\left(
\begin{array}{c}
[\epsilon_2+m_2]\eta({\bf p'})-
|{\bf p}|[\hat{\bf p}\cdot\hat{\bf p'}+
i\hat{\bf p}\times\hat{\bf p'}\cdot\vec{\sigma}]\zeta({\bf p'}),
\\
\sigma\cdot\hat{\bf p}\left[-|{\bf p}|\eta({\bf p'})+
[\epsilon_2-m_2][\hat{\bf p}\cdot\hat{\bf p'}+
i\hat{\bf p}\times\hat{\bf p'}\cdot\vec{\sigma}]\zeta({\bf p'})\right]
\end{array}\right)\chi,
\end{eqnarray}
and
\begin{eqnarray}
\Lambda_{-}(-{\bf p})\phi({\bf p'})=\frac{1}{2\epsilon_2}
\left(
\begin{array}{c}
[\epsilon_2-m_2]\eta({\bf p'})
+|{\bf p}|[\hat{\bf p}\cdot\hat{\bf p'}+
i\hat{\bf p}\times\hat{\bf p'}\cdot\vec{\sigma}]\zeta({\bf p'})
\\
\sigma\cdot\hat{\bf p}\left[ |{\bf p}|\eta({\bf p'})+
[\epsilon_2+m_2][\hat{\bf p}\cdot\hat{\bf p'}+
i\hat{\bf p}\times\hat{\bf p'}\cdot\vec{\sigma}]\zeta({\bf p'})\right]
\end{array}\right)\chi.
\end{eqnarray}
We have taken $g^2_{Dq}=g^2_{dqq}e^{-\lambda^2 {\bf q}^2}$ where ${\bf q}={\bf p}+{\bf p}'$ and
$\lambda=0.18\mbox{fm}^{-2}$, $g_{dqq}=14.15$\cite{Keiner:1995bu}, 
$m_1=650$ MeV and $m_2=350$ MeV for the diquark and quark, respectively.

\section{MIT bag model for nucleon}
The nucleon is assumed as MIT bag of the confined 
quarks\cite{Chodos:1974je,Guichon:1987jp,Zakout:1998gs}.
The quark field $\psi_{q}(\vec{r},t)$ inside the nucleon bag of radius $R$
satisfies the Dirac equation
\begin{eqnarray}
\left[ i\gamma^{\mu}\partial_{\mu}-m_{q}^{0} \right]\psi_{q}(\vec{r},t)=0,
\label{Dirac} 
\end{eqnarray}
where $m_{q}^{0}$ is the current mass of a quark of flavor $q$. 
The current quark masses are taken $m_u=m_d=0$ for the up and down flavor quarks.
The single-particle quark energy is given by
\begin{eqnarray}
\epsilon_q=\Omega_q/R
\label{EPN} 
\end{eqnarray}
where
\begin{eqnarray}
\Omega_q=\sqrt{ x^{2}_q + R^{2}{m}^{2}_{q} }.
\label{Omegnk} 
\end{eqnarray}
For a given value of the bag radius $R$, the quark momentum
$x_{q}$ is  determined by the confinement boundary condition 
at the bag surface which,
for quarks of flavor $q$ in a spherical bag, reduces to
$j_0({x_{q}})=\beta_{q} j_1({x_{q}})$, where
$\beta_{q}=\sqrt{\frac{{\Omega_q}-R m_q }{{\Omega_q} + R m_q }}$
and $j_0,j_1$ are spherical Bessel functions 
of order zero and one, respectively.
The bag energy is given by
\begin{eqnarray}
E_{\mbox{bag}}= \sum^{n_q}_q \frac{\Omega_q}{R}
- \frac{Z}{R}+\frac{4\pi}{3}R^{3}  B,
\end{eqnarray}
where $\sum^{n_q}_q \frac{\Omega_q}{R}$
is the total constituent quarks kinetic energy inside the bag
and $\frac{Z}{R}$ term is the zero-point energy for quarks
and $B$ is the bag parameter.
The spurious center-of-mass energy is subtracted
to obtain nucleon mass,
\begin{eqnarray}
M^{*}=\sqrt{{E_{\mbox{bag}}}^2 - <p^{2}_{\mbox{cm}}>},
\label{MNSTAR}          
\end{eqnarray}
where
\begin{eqnarray}
<p^{2}_{\mbox{cm}}>=
\sum^{n_q}_q x_{q}^2 / R^2.
\label{PCM}
\end{eqnarray}
The bag radius $R$ is obtained through the minimization of the baryon mass with
respect to its bag radius
\begin{eqnarray}
\frac{\partial M^{*}}{\partial R}=0.
\label{MNR}
\end{eqnarray}
The bag parameter $B$ is taken as $B_{0}$ corresponding 
to its value for a free nucleon.
The values $B^{1/4}_{0}=188.1$ MeV and $Z_{0}=2.03$
are chosen to reproduce the free nucleon mass
$M_{N}$ at its experimental
value 939 MeV and bag radius $R_{0}=0.60$ fm.
The current quark mass $m_{q}$ is taken equal to zero.

In order to consider the thermal effects for the nucleon mass $M(T)$, 
we assume that the bag parameter dissolves with temperature 
as follows
\begin{eqnarray}
B\equiv B(T)=B_0\times {\cal R}^{4\delta}(T)
\end{eqnarray}
where 
\begin{eqnarray}
{\cal R}(T)=\left(1-b\frac{T^2}{T^2_c}\right)^{1/2}.
\end{eqnarray}
With a choice $\delta=1$, the nucleon mass at finite temperature 
reduces to
\begin{eqnarray}
M_N(T)=M_N(0)\times {\cal R}(T).
\end{eqnarray}
This choice is relevant to the thermal dissociation of the linear confinement
$\sigma(T)=\sigma_0 {\cal R}(T)$
\cite{Pisarski:1982cn,Kaczmarek:1999mm}
and its possible connection to the bag pressure dissociation
$B^{1/4}(T)=B_0^{1/4} {\cal R}(T)$. This connection leads to the choice $\delta=1$. 
\section{Results and Conclusions}
We have studied the nucleon as a bound state of a scalar-diquark and a quark 
by diagonalizing the BSE at finite temperature. 
The interaction is taken as a quark exchange in the adiabatic approximation.
The quark and scalar-diquark substructures and their modifications 
with momentum and temperature are introduced self-consistently in the BSE. 
The interaction modification with temperature 
is approximated using a mixed representation of the imaginary time formalism in the propagator 
and then the adiabatic (instantaneous) approximation is adopted by substituting the time component 
to zero $q_0=0$ after evaluating the Matsubara frequency sums not before.
We have included explicitly in our calculations the smooth decreasing 
of the constituent quark and diquark masses with temperature. 
The thermal variations of the constituent particle masses are based on the lattice-QCD calculation.
In Fig.(I), we display the dependence of constituent quark 
and scalar-diquark masses on their substructure momenta at zero temperature.
The quark and diquark acquire their constituent masses 350 and 650 MeV, respectively, at zero momentum.
They decrease with momentum to reach their current masses at large momentum.
The diquark acts as a weakly bound state of quark-quark and its mass varies 
with temperature in the same order of variation of its constituent quarks.
In Fig. (II), we display the quark and diquark masses versus temperature at zero momentum.
It is shown that the constituent masses decrease with temperature as ${\cal R}(T)$.
The thick lines correspond $p_4=\pi T$, while the thin lines correspond $p_4=0$.
It is shown that the first order Matsubara frequency correction has a very small
correction to the adiabatic approximation.

We have calculated the nucleon mass by diagonalizing BSE with several temperatures. 
The nucleon mass versus temperature is displayed in Fig.(III). 
We have found that $2M_N(T)$ is always less than $3m_{diq}(T)$ 
until the temperature reaches just below the critical one. 
This means that the nucleon bound state persists to exist until 
the temperature reaches a critical one.
Therefore, the quarks continue to confine  
in the hadronic phase and they shall not deconfine spontaneously 
until the temperature reaches the critical one. 
Furthermore, we have found just below  the critical temperature, 
the nucleon bound state ceases to exist. 
It deconfines to its constituents spontaneously at the critical temperature. 
In our calculations, the critical temperature is taken $T_c=170$ MeV.
This probably a signature for the first order phase transition. 
Furthermore, the nucleon mass decreases smoothly with temperature 
in the same manner the constituent quark mass decreases with temperature 
and at the same order of the linear confinement dissociation
in the lattice-QCD calculation. 
These results are compared with those for MIT bag model. 
We assume that MIT bag pressure is temperature dependent 
and it dissociates as $B^{1/4}(T)=B_0^{1/4} {\cal R}(T)$ and $\delta=1$.  
This dissociation is at the same order of the linear confinement dissociation 
in the potential models and lattice-QCD calculations\cite{Pisarski:1982cn,Kaczmarek:1999mm}.
We have found with this choice, the nucleon mass decreases 
smoothly with temperature as $M_N(T)/M_N(0)= {\cal R}(T)$. 
Furthermore, the choice $\delta=1/4$ doesn't match the mass spectrum for BSE.

Furthermore, the solutions of BSE and MIT bag are in a good agreement with each other 
but the MIT bag model lacks the first order phase transition 
that found for BSE in the critical temperature.
However, the MIT bags might have the first order phase transition 
when they are embedded in a nuclear matter because of their interaction 
with the medium and the the production of particles and antiparticles. 
Finally, in our conclusions, we argue the nucleon mass in a hot bath 
should be a temperature dependent and it decreases smoothly
as $M_N(T)=M_N{\cal R}(T)$ where ${\cal R}(T)=\left(1-b\frac{T^2}{T^2_c}\right)^{1/2}$.

\begin{figure}
\includegraphics{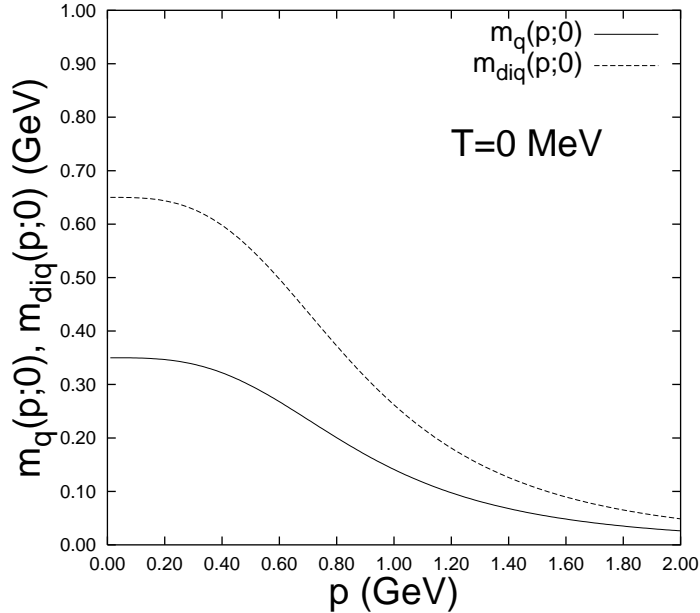}%
\caption{\label{fig1}
The constituent quark and scalar-diquark masses
versus momentum at zero temperature.}
\end{figure}
\begin{figure}
\includegraphics{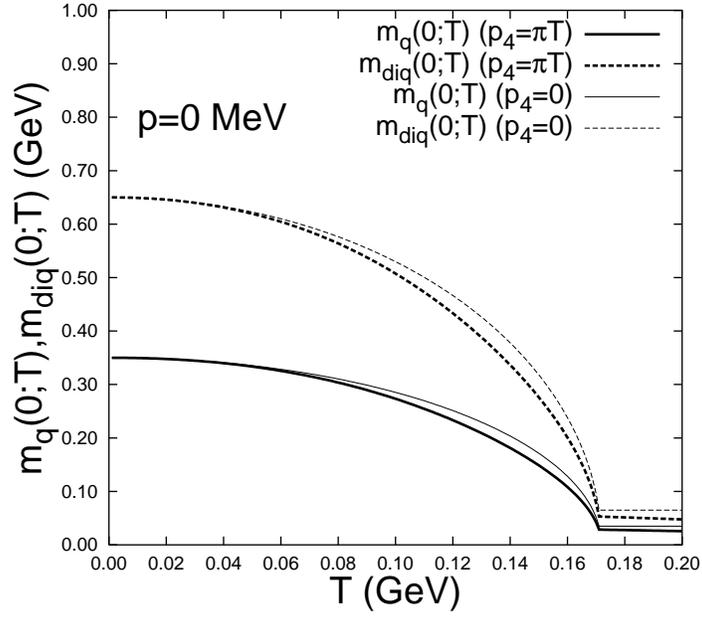}%
\caption{\label{fig2}
The constituent quark and scalar-diquark masses with ${\bf p}=0$
versus temperature.}
\end{figure}
\begin{figure}
\includegraphics{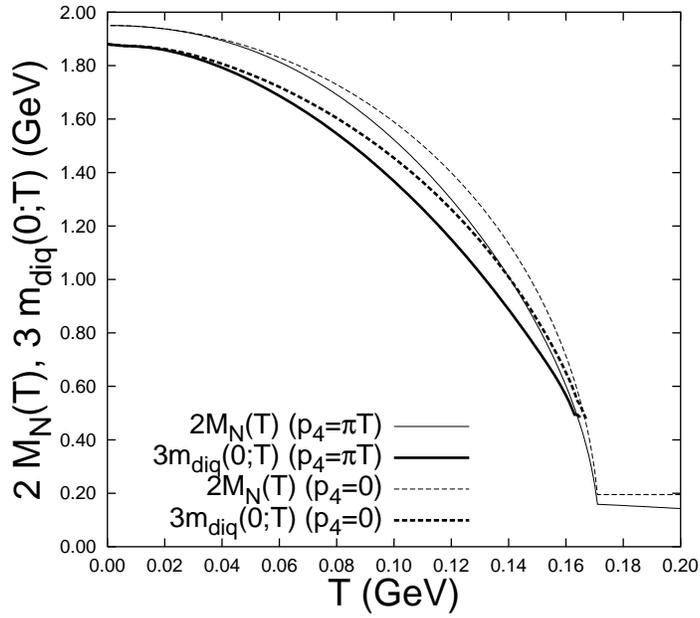}%
\caption{\label{fig3}
Nucleon mass calculated using BSE and the scalar-diquark constituent mass
at ${\bf p}=0$ versus temperature.
It is shown that the bound state persists to exist
until the temperature reaches just below the critical one
and then the bound state ceases to exist at the critical one.}
\end{figure}
\begin{figure}
\includegraphics{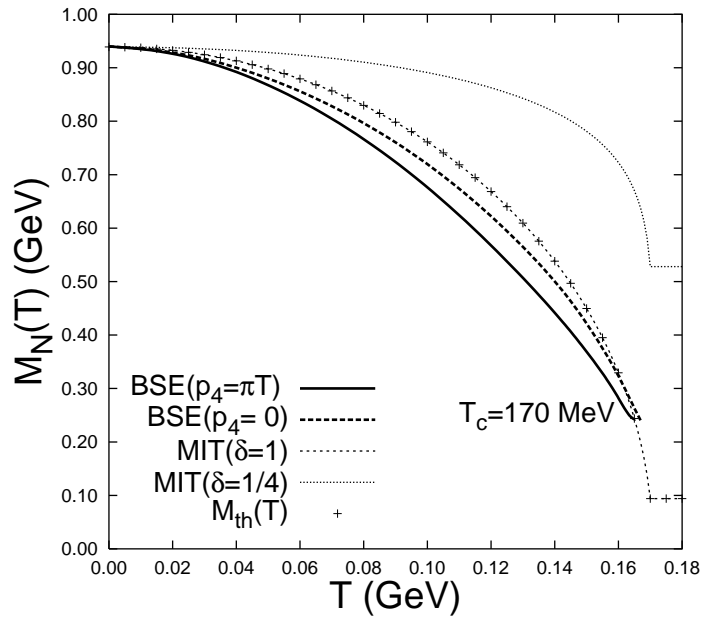}%
\caption{\label{fig4}
The nucleon mass $M_N(T)$ versus temperature. The nucleon mass is obtained by
BSE and MIT bag models with several sets as explained in the text.}
\end{figure}

\begin{acknowledgments}
I. Z. gratefully acknowledges support from the Alexander von Humboldt Foundation.
He thanks H. R. Jaqaman, C. Greiner,  J. Schaffner-Bielich and R. Sever for the discussions and comments. 
\end{acknowledgments}


\begin{thebibliography}{99}
\bibitem{Adami:1990sv}
C.~Adami, T.~Hatsuda and I.~Zahed,
Phys.\ Rev.\ D {\bf 43}, 921 (1991).

\bibitem{Adami:1992at}
C.~Adami and I.~Zahed,
Phys.\ Rev.\ D {\bf 45}, 4312 (1992).

\bibitem{Kacir:1995gy}
M.~Kacir and I.~Zahed,
Phys.\ Rev.\ D {\bf 54}, 5536 (1996)

\bibitem{Ropke:1986mh}
G.~Ropke, D.~Blaschke and H.~Schulz,
Phys.\ Lett.\ B {\bf 174} (1986) 5.
                                                                                

\bibitem{Keiner:1995bu}
V.~Keiner,
Z.\ Phys.\ A {\bf 354}, 87 (1996)

\bibitem{Keiner:1996at}
V.~Keiner,
Phys.\ Rev.\ C {\bf 54}, 3232 (1996)

\bibitem{Hellstern:1997pg}
G.~Hellstern, R.~Alkofer, M.~Oettel and H.~Reinhardt,
Nucl.\ Phys.\ A {\bf 627}, 679 (1997)
\bibitem{Oettel:2000jj}
M.~Oettel, R.~Alkofer and L.~von Smekal,
Eur.\ Phys.\ J.\ A {\bf 8}, 553 (2000)
\bibitem{Alkofer:2004yf}
R.~Alkofer, A.~Hoell, M.~Kloker, A.~Krassnigg and C.~D.~Roberts,
[arXiv:nucl-th/0412046].

\bibitem{Blaschke:1998gk}
D.~Blaschke, Y.~L.~Kalinovsky and P.~C.~Tandy,
[arXiv:hep-ph/9811476].
\bibitem{Blaschke:2000gd}
D.~Blaschke, G.~Burau, Y.~L.~Kalinovsky, P.~Maris and P.~C.~Tandy,
Int.\ J.\ Mod.\ Phys.\ A {\bf 16}, 2267 (2001)

\bibitem{Roberts:2000aa}
C.~D.~Roberts and S.~M.~Schmidt,
Prog.\ Part.\ Nucl.\ Phys.\  {\bf 45}, S1 (2000)


\bibitem{Ikeda:2001vc}
T.~Ikeda,
Prog.\ Theor.\ Phys.\  {\bf 107}, 403 (2002)

\bibitem{Iida:2003gq}
H.~Iida, M.~Oka and H.~Suganuma,
Nucl.\ Phys.\ Proc.\ Suppl.\  {\bf 129} (2004) 602

\bibitem{Iida:2004ih}
H.~Iida, M.~Oka and H.~Suganuma,
[arXiv:hep-ph/0410222].

\bibitem{Chodos:1974je}
A.~Chodos, R.~L.~Jaffe, K.~Johnson, C.~B.~Thorn and V.~F.~Weisskopf,
Phys.\ Rev.\ D {\bf 9}, 3471 (1974).

\bibitem{Guichon:1987jp}
P.~A.~M.~Guichon,
Phys.\ Lett.\ B {\bf 200}, 235 (1988).

\bibitem{Zakout:1998gs}
I.~Zakout and H.~R.~Jaqaman,
Phys.\ Rev.\ C {\bf 59}, 962 (1999)

\bibitem{Pisarski:1982cn}
R.~D.~Pisarski and O.~Alvarez,
Phys.\ Rev.\ D {\bf 26}, 3735 (1982).

\bibitem{Kaczmarek:1999mm}
O.~Kaczmarek, F.~Karsch, E.~Laermann and M.~Lutgemeier,
Phys.\ Rev.\ D {\bf 62}, 034021 (2000)


\end{thebibliography}
\end{document}